 \documentclass[final,3p,times]{elsarticle}

\usepackage{amssymb}
\usepackage{amsmath}

\journal{Chaos, Solitons \& Fractals}

\usepackage[T1]{fontenc}
\usepackage[utf8]{inputenc}
\usepackage{lmodern}
\usepackage{hyperref}
\usepackage{multirow}
\usepackage{xcolor}
\usepackage{mathtools} 
\usepackage[caption=false,labelformat=simple]{subfig}
\biboptions{sort&compress}
\begin{document}

\begin{frontmatter}



\title{Unified Framework for Binary-Choice Dynamics: Analysis and Applications}

\author[pwr]{Arkadiusz J\k{e}drzejewski} 
\ead{arkadiusz.jedrzejewski@pwr.edu.pl}
\affiliation[pwr]{organization={Department of Computational Social Science, Wroc\l{}aw University of Science and Technology}, 
postcode={50-370},
state={Wroc\l{}aw}, 
country={Poland}}

\author[ua]{Jos\'{e} F. F. Mendes} 

\affiliation[ua]{organization={Department of Physics \& I3N, University of Aveiro},
postcode={3810-193}, 
state={Aveiro},
country={Portugal}}

\begin{abstract}
We demonstrate how the unified framework for binary-choice dynamics can be used to study the role of annealed and quenched disorders in homogeneous and heterogeneous systems.
The framework defines the structure of interactions between agents without imposing their functional forms.
Such a high level of generality allows us to connect many different models across disciplines and find universal rules that apply to all of them.
Within this framework, agents update their states under the influence of two competing mechanisms chosen according to individual preferences.
We review the literature to classify existing models as homogeneous or heterogeneous based on their preference distribution, and we discuss the role of annealed (changing) and quenched (fixed) disorders in modeling these preferences.
Using the framework, we derive a constraint on the transition rates. When a model meets this condition, three major things happen: annealed and quenched dynamics become equivalent, any heterogeneous system can be mapped into a homogeneous one, and oscillations cannot emerge. 
We illustrate these consequences using models from statistical physics, opinion dynamics, and disease spreading.
Finally, we discuss the framework limitations and its potential further developments.

\end{abstract}



\end{frontmatter}


\section{Introduction}
Across diverse domains, from statistical physics to social science, binary-choice models are widely used to investigate how interactions between individual agents lead to collective outcomes \cite{Sta:etal:26, Gle:13}.
In most cases, such models are constructed by specifying particular functions that describe how agents influence each other.
While this approach has produced many important insights, it also has some limitations. 
The findings from the analysis of such models remain model-specific, and a modification of the interaction function may drastically change the model behavior and the conclusions.
As a result, it is difficult to identify general principles that apply across different models, and even more so across different fields.

Interestingly, despite their diversity, many binary-choice models share a common underlying structure: agents change their states under the influence of two competing mechanisms chosen according to individual preferences.
Depending on a specific model, these mechanisms can represent different physical \cite{Gar:Lab:Mar:87,Tom:Oli:San:91,Tam:Ale:Gup:94} or biological \cite{Kee:Eam:05,Pag:Now:02} processes, economic effects \cite{Byr:etal:16}, or social behaviors \cite{Jed:Szn:19,Yan:etal:21}.
In a recent Letter \cite{Jed:Men:25}, we use this structure to formulate a unified framework that captures a broad class of models. 
Instead of imposing specific forms on the transition rates defining the mechanisms, we leave them as unspecified functions. 
Although it may seem that such a framework is too general to bring any useful insight, it allows us to derive universal conclusions that remain valid regardless of the model details. 
This approach is conceptually similar to generalized modeling \cite{Gro:Feu:06,Mas:Gro:22}, where dynamical systems are constructed by postulating the structure of interactions rather than their precise forms and then  analyzed using local bifurcation theory \cite{Kue:Sie:Gro:13}.

A different approach to relating binary-state models was proposed by Galam, where diverse models of opinion dynamics are mapped into a sequential probabilistic model \cite{Gal:05,Gal:22}. 
The core idea is to cast a specific update rule into a probabilistic formula.
Then, the discrete-time evolution equation is derived through  successive local updates of this formula interspersed with repeated reshuffling of agents. By ignoring special arrangement, this procedure effectively simulates a well-mixed population of homogeneous agents.

A major advantage of our framework is its versatility that allows for a direct comparison of commonly used modeling approaches. Based on the distribution of individual preferences, the framework can describe homogeneous systems, where all agents share the same preference, or heterogeneous ones, where preferences vary among them \cite{Yan:etal:21,Jed:Her:24}.
Furthermore, it distinguishes between annealed dynamics, where agents change their preferences over time, and quenched dynamics, where preferences remain fixed once assigned \cite{Szn:Szw:Wer:14,Jed:Szn:20}.
These different modeling assumptions often lead to mixed results, making it unclear why some models are highly sensitive to the choice of dynamics or heterogeneity while others are completely indifferent \cite{Szn:Szw:Wer:14,Jed:Szn:17,Yan:etal:21,Jed:Her:24,Pra:Mul:Sen:26}.
The framework solves this puzzle by identifying a simple balancing condition on the transition rates \cite{Jed:Men:25}.
When this condition is satisfied, it leads to three major consequences: (1) Any heterogeneous system can be mapped to a homogeneous one, (2) the annealed and quenched dynamics become equivalent, and (3) the system cannot exhibit any oscillatory behavior. 

In this article, our main goal is to demonstrate the practical applications of the framework beyond simply unifying existing models. 
Based on examples from opinion dynamics, statistical physics, and epidemiology, we show how different specific models fit into the framework, which of them satisfy the balancing condition, and how this impacts their behavior.
Although the core framework was introduced in Ref.~\cite{Jed:Men:25}, we preset here the step-by-step mathematical derivations for completeness.
We also include a short review of how both dynamics and heterogeneity have been treated in the literature to provide a clear context for our results.
Finally, we discuss the limitations of the framework and directions for its future development.

The remainder of the paper is organized as follows: Section~\ref{sec:model} defines the unified framework.
Sections~\ref{sec:dyn} through \ref{sec:equi} provide a literature review on the foundational concepts:
Section~\ref{sec:dyn} discusses the annealed and quenched dynamics;
Section~\ref{sec:homo}, homogeneous and heterogeneous populations;
and Section~\ref{sec:equi}, the model-dependent nature of the equivalence between these dynamics.
Section~\ref{sec:anal} contains a detailed mathematical analysis and the derivation of the balancing condition.
Section~\ref{sec:dis} focuses on the cross-disciplinary  applications of the framework, while Section~\ref{sec:limit} examines its limitations and possible extensions.
Finally, Section~\ref{sec:con} concludes the paper.

\begin{figure}[!t]
	\centering
	\includegraphics[width=\linewidth]{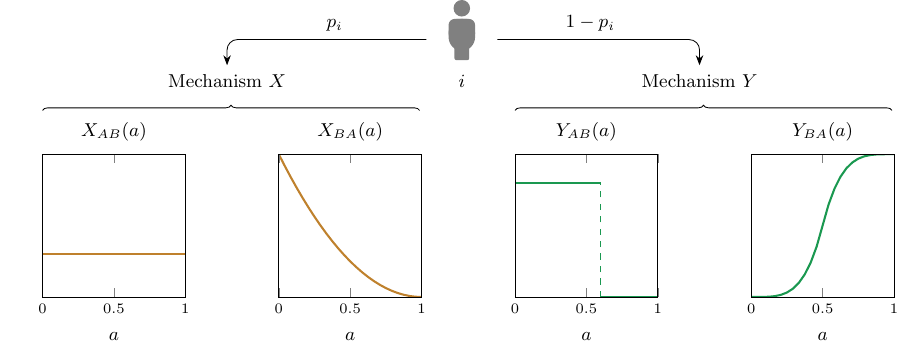}
	\caption{\label{fig:model-vis} Agent $i$ chooses between two options, $A$ and $B$, under the influence of either mechanism $X$ or $Y$. The mechanisms are defined by transition probabilities (or rates), which specify the likelihood of switching from $A$ to $B$ ($AB$) and from $B$ to $A$ ($BA$).
    These functions depend on the current fraction of agents choosing option $A$ in the system, denoted by $a$, and they may take arbitrary forms.
    Agent $i$ follows mechanism $X$ with probability $p_i$; otherwise, it uses mechanism $Y$ to update its state. 
 }
\end{figure}
\section{Modeling Framework}
\label{sec:model}
Let us consider a population of agents where every individual has the same chance to interact with anyone else.
In this context, the term agent is used in a broad sense.
It may refer to a person expressing an opinion or making a decision \cite{Prz:Szn:Wer:14, Yan:etal:21,Lui:Che:26}, an individual exposed to infection \cite{Kee:Eam:05}, a biological entity in evolutionary dynamics \cite{Pag:Now:02}, or a physical concept, such as a spin in the Ising model \cite{Gar:Lab:Mar:87,Tom:Oli:San:91,Tam:Ale:Gup:94}. 
An agent can also represent a collective unit, like a group of consumers or a household \cite{Wer:Kow:Wer:18,Kow:etal:14}.
Such a  well-mixed population can be represented by a complete graph where every pair of nodes is connected. 
Nodes are agents, and links represent possible interactions.

Each agent can be in one of two states, $A$ or $B$.
Their interpretation depends on the context. 
In social dynamics, these states may represent different opinions or choices \cite{Jed:Szn:19,Yan:etal:21}.
In physical systems, they correspond to spin orientations \cite{Gar:Lab:Mar:87,Tom:Oli:San:91,Tam:Ale:Gup:94}, while in epidemiology, they may indicate the health status, such as susceptible or infected \cite{Kee:Eam:05,Bra:08,Bar:Bar:Ves:08}.

The evolution of states occurs through two mechanisms, $X$ and $Y$, which define the transition probabilities (or rates) for an agent to switch states. 
We do not impose any specific functional forms on them. 
We only assume that they depend on the current state of the system, characterized by the total fraction of agents in state $A$, denoted by $a$. 
Let $X_{AB}(a)$ and $X_{BA}(a)$ represent the probabilities that an agent switches from state $A$ to $B$, and $B$ to $A$, respectably, through mechanism $X$.
Similarly, $Y_{AB}(a)$ and $Y_{BA}(a)$ denote the corresponding transition probabilities for mechanism $Y$.
Let us note that these mechanisms can be also defined in terms of transition rates instead of probabilities.
While a rate-based formation is more general, the probability-based approach may be more intuitive.

The system evolves through a random sequential updating scheme.
At each time step, a single agent $i$ is chosen uniformly at random.
This agent follows mechanism $X$ with probability $p_i$, or mechanism $Y$ with probability $1-p_i$.
In the social context, the parameter $p_i$ can be interpreted as the individual preference of agent $i$ for mechanism $X$ or its personal trait. In more abstract cases, it may just represent the relative frequency of the two mechanisms. 
A schematic overview of the updating rules is provided in Fig.~\ref{fig:model-vis}.

\begin{figure}[!b]
	\subfloat{\label{fig:que-ann-vis:a}}
    \subfloat{\label{fig:que-ann-vis:b}}
	\centering
	\includegraphics[width=\linewidth]{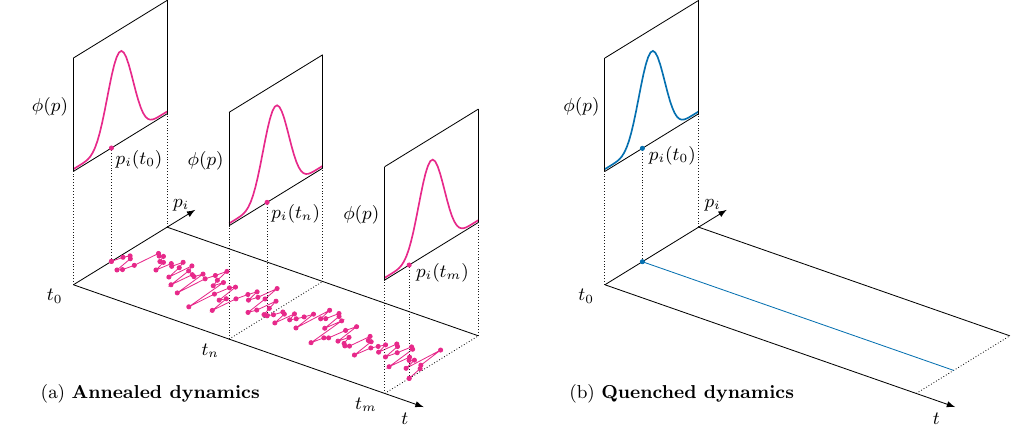}
	\caption{\label{fig:que-ann-vis} (a) Annealed dynamics: At every time step $t$, agent $i$ receives a new preference for mechanism $X$, denoted by $p_i(t)$. The highlighted moments, $t_0$, $t_n$, $t_m$, in the plot show examples of these updates. Each new value of $p_i$ is drawn independently from the same distribution $\phi(p)$, for every agent and at every time step.
    (b) Quenched dynamics: At the start of the process, $t_0$, agent $i$ is assigned preference $p_i(t_0)$. This value is drawn independently from the same distribution $\phi(p)$ for each agent, and it remains fixed for all future times.
 }
\end{figure}

The parameters $p_i$ are drawn independently for all agents from the same distribution $\phi(p)$, which may be discrete or continuous. 
Depending on the chosen dynamics, annealed or quenched, $p_i$ may either remain fixed or change over time. 
However, in both cases, the distribution $\phi(p)$ itself remains unchanged over time.
These two approaches are implemented as follows:
\begin{itemize}
\item Annealed dynamics: At each update step, each agent $i$ redraws its $p_i$ from the distribution $\phi(p)$. The value used in one step does not influence the value selected in the next step, see Fig.~\ref{fig:que-ann-vis:a}.

\item Quenched dynamics: Each agent $i$ is assigned a specific $p_i$ drawn from the distribution $\phi(p)$ at the beginning of the process. This value is a permanent attribute of the agent and remains fixed throughout the entire system evolution, see Fig~\ref{fig:que-ann-vis:b}.

\end{itemize}

Despite different microscopic implementations, as the number of agents increases, the empirical distribution of preferences in the population converges to the underlying distribution $\phi(p)$ in both cases.

\section{Annealed and Quenched Dynamics}
\label{sec:dyn}
In modeling complex systems, annealed and quenched dynamics capture two different time scales of change.
The distinction lies in how fast some individual features, like agents' preferences in our case, evolve compared to the system macro-level behavior \cite{Szn:Szw:Wer:14,Jed:Szn:17,Jed:Szn:20}.

\subsection{Annealed Dynamics}
Annealed dynamics describes systems where random features vary as the system progresses.
This approach is suited for modeling situational or environmental fluctuations.
In physics, a classic example is the thermal fluctuations of atoms in a crystal.
In social systems, annealed dynamics models highly adaptive behavior where individuals change their inclinations based on varying external factors, such as mood, social context, or levels of uncertainty \cite{Ken:etal:18, Mor:etal:12, Hop:Lal:13}.
This aligns with a situation-oriented perspective in psychology, where human behavior is viewed as a response to changing situational factors rather than a manifestation of stable dispositions \cite{Don:Luc:Fle:09}.
For instance, the use of different learning strategies is often a transient response to contextual nuances rather than a fixed trait \cite{Ren:etal:11,Ken:etal:18,Mce:etal:08}.
Similarly, in epidemiological models, annealed dynamics can represent a time-varying exposure to infection due to environmental changes.

\subsection{Quenched Dynamics}
Quenched dynamics describes systems where random features are fixed through time evolution.
In physics, this concept can be traced back to the studies of spin glasses, disordered magnetic materials where local impurities are permanent \cite{Ste:New:13}.
To study these systems, physicists introduced fixed random variables into the classic Ising model, which was later adapted also to describe social phenomena. 
In these social applications, a quenched random field represents agents' personal inclinations towards one opinion \cite{Gal:Mos:91,Gal:97}.
In a social context, quenched dynamics represents persistent individual traits or dispositions, so it resonates with the personality-oriented perspective in psychology, which posits that people are characterized by a set of stable and enduring qualities that make their behavior persistent and predictable \cite{Don:Luc:Fle:09}.
For example, some individuals are inherently more risk-prone than others or possess stable social ranks that dictate their behavior over long periods \cite{Ken:etal:18,Eff:etal:08}.

\section{Homogeneous and Heterogeneous Systems}
\label{sec:homo}
\begin{figure}[!t]
	\subfloat{\label{fig:distributions:a}}
    \subfloat{\label{fig:distributions:b}}
    \subfloat{\label{fig:distributions:c}}
    \subfloat{\label{fig:distributions:d}}
	\centering
	\includegraphics[width=\linewidth]{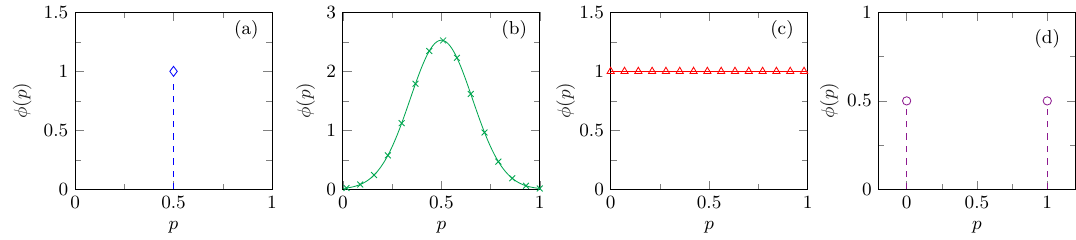}
	\caption{\label{fig:distributions} Illustration of the preference distributions with the same mean preference $\bar{p}=0.5$: (a) a degenerate distribution, (b)  a Bernoulli distribution, (c) a uniform distribution on $[0,1]$, and (d) a normal distribution centered at $p=0.5$ with variance $\sigma^2=1/40$. The degenerate distribution models a homogeneous population where all agents have the same preference, whereas the three other distributions model a heterogeneous population where agents differ in their preferences.
 }
\end{figure}

The distinction between homogeneous and heterogeneous systems can be considered at different levels. In general, agents are characterized by different parameters. If a given parameter takes a single value for all agents, the system is homogeneous. In contrast, if a parameter varies across the population, the system is heterogeneous.
In our framework, whether the system is heterogeneous depends on the distribution of preferences, $\phi(p)$.
If $\phi(p)$ is degenerate, i.e., $p_i$ takes only one value, the system is homogeneous, see Fig.~\ref{fig:distributions:a}. Otherwise, the system is heterogeneous, as illustrated by the examples in Figs.~\ref{fig:distributions:b}-\ref{fig:distributions:d}.

\begin{figure}[!t]
	\centering
	\includegraphics[width=0.9\linewidth]{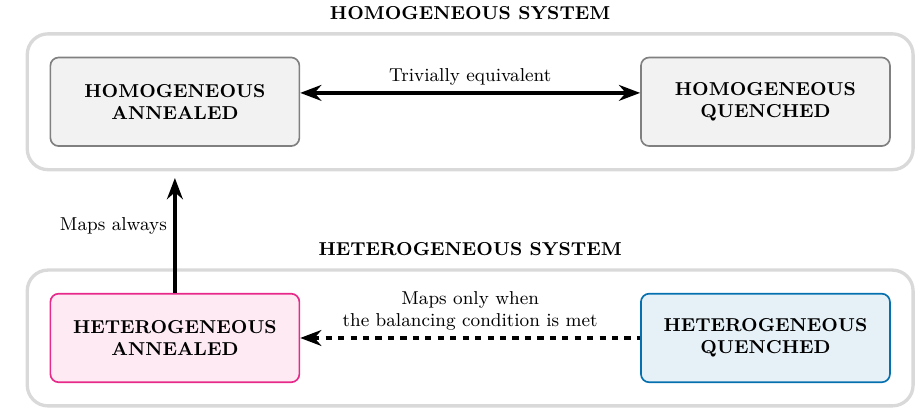}
	\caption{\label{fig:diagram} For homogeneous systems, the annealed and quenched dynamics are trivially equivalent since only a single value of preference is possible (top container). For heterogeneous systems, however, the two dynamics generally lead to different results, and only under the balancing condition can the quenched dynamics be mapped into the annealed one (bottom container). Moreover, in the annealed case, only the mean of the distribution matters, so the dynamics of a heterogeneous population reduces to that of a homogeneous population with a preference equal to the distribution mean (vertical arrow). Consequently, when the balancing condition is satisfied, any heterogeneous system can be mapped into a homogeneous one. 
	}
\end{figure}

This distinction is important from a modeling standpoint as annealed and quenched dynamics are trivially equivalent only within homogeneous systems.
For heterogeneous ones, this equivalence is far from obvious, and as we will show, it holds only under a very special condition.
Crucially, the equivalence implies that such a heterogeneous system can be mapped to a homogeneous one, which simplifies mathematical analysis and rules out any oscillatory behavior.
Figure~\ref{fig:diagram} illustrates this interplay between dynamics and system heterogeneity.

\subsection{Homogeneous Systems}
In kinetic Ising models with competing dynamics, different transition rules occur with some probabilities that are the same for all spins \cite{Gon:Gar:Leb:87,Tom:Oli:89,Dum:God:24}.
Although theoretically a system may involve multiple competing mechanisms \cite{Gar:Lab:Mar:87}, it is often limited to two.
These typically fall into one of two categories.
The first involves competing temperatures, where the system evolves under a single type of dynamics, such as Glauber or Metropolis, but under a fluctuating environment that randomly switches the temperature at each time step \cite{Gar:Lab:Mar:87,Tom:Oli:San:91,Tam:Ale:Gup:94}.
The second category consists of competing different microscopic processes, a spin-flip dynamics, which simulates contact with a heat bath, and spin-exchange Kawasaki dynamics \cite{Gon:Gar:Leb:87,Szo:00,Tom:Oli:89,Dum:God:24,Dum:God:23}, which simulates energy flux into the system.
This type of completion has also been studied in the Heisenberg model \cite{Dum:Cos:God:25}.

In homogeneous social systems, all agents share the same probability to act in a specific way. 
For example, in the majority-vote model, an agent adopts the majority opinion with a fixed probability and opposes it otherwise \cite{Oli:92,Cro:Tom:05}.
Similar formulations appear across various models of opinion dynamics, indulging the Sznajd model \cite{Sch:04,Lam:lop:Wio:05}, the threshold model  \cite{Now:Szn:19,Now:Gra:Szn:22,Now:Szn:20}, or the nonlinear voter model \cite{Cas:Mun:Pas:09,Per:etal:18,Mus:etal:25}, including versions with thresholds  \cite{Nyc:etal:18,Vie:Cel:18,Vie:etal:20},  anticonformity \cite{Nyc:Szn:13,Nyc:Szn:Cis:12}, or biases \cite{Per:etal:21,Don:etal:25,Pra:Mul:Sen:26}.

\subsection{Heterogeneous Systems}
When it comes to modeling heterogeneous systems, researcher often capture diversity in a simplified coarse-grained manner. 
Commonly, the agents are divided into just two groups with different behavioral rules.
This effectively creates a Bernoulli distribution of preferences, see Fig.~\ref{fig:distributions:b}.
For example, in the Galam model, a fraction of the population is randomly chosen to be contrarians, while the rest are conformists \cite{Gal:04,Gal:26,Bor:Gal:06}.
Such division into agents with specific behaviors are present in the Sznajd model \cite{Sch:04}, the majority-vote model \cite{Vil:Mor:Sou:12,Vil:Sou:17,Jav:14,Kra:18,Vil:etal:19,Gra:etal:22,Oli:etal:24}, the nonlinear $q$-voter model \cite{Jed:Szn:20,Jed:Szn:17,Jed:Szn:22,Pra:Mul:Sen:26,Jav:Squ:15,Now:Sto:Szn:21,Now:Szn:22}, the threshold model \cite{Gra:Li:20,Juu:Por:19}, or the voter model \cite{Cza:etal:22}.
In these contexts, agents can be assigned identities such as inflexibles \cite{Gal:Jac:07} or zealots \cite{Mob:03} who favor one opinion, contrarians or anticonformists who oppose the group influence \cite{Sta:Mar:04,Jed:Szn:17,Now:Szn:22}, or independent individuals who ignore social pressure entirely \cite{Szn:Szw:Wer:14,Now:Sto:Szn:21}.
A similar split appears in decision-making models, where agents are divided into individual and social learners \cite{Yan:etal:21,Jed:Her:24,Wu:etal:25}.  

More broadly heterogeneous cases are rarely considered, even though some  studies have explicitly noted that such extensions are possible \cite{Vil:Mor:Sou:12,Jed:Her:24}.
For example, in a decision-making dynamical system model, researchers moved beyond the standard Bernoulli distribution to test uniform, truncated normal, and beta distributions  of learning preferences \cite{Yan:etal:21}.
Interestingly, while the uniform and truncated normal led to identical system behavior, the beta distribution yielded completely different results.
This illustrates that the specific choice of distribution can change the outcome, so it should be selected thoughtfully.
At the same time, it remains unclear why some changes in the distribution impact the behavior of the system while others do not.

\section{Annealed and Quenched Dynamics in Heterogeneous Systems}
\label{sec:equi}
A fundamental question that naturally arises is whether annealed and quenched dynamics lead to the same macroscopic behavior for heterogeneous systems.
Since any heterogeneous system under annealed dynamics is equivalent to a homogeneous system where all agents have the same preference equal to the mean of the heterogeneous distribution, $\bar{p}=\int p\phi(p)dp$, it is enough to compare the quenched heterogeneous system with such a homogeneous system, see Fig.~\ref{fig:diagram}.

This problem has been explored across various specific models.
However, most studies divide the population into two distinct behavioral groups, which corresponds to Bernoulli distributed preferences \cite{Szn:Szw:Wer:14, Sta:Mar:04, Pra:Mul:Sen:26,Sch:04,Jed:Szn:17,Jed:Szn:22,Jed:Szn:20,Byr:etal:16,Jed:Her:24}.
This means that under quenched dynamics, a fixed fraction $\bar{p}$ of the population permanently adopts one behavior (e.g., anticonformity), while the remaining fraction $1-\bar{p}$ adopts another (e.g., conformity).
On the other hand, annealed dynamics describes a homogeneous population where the behavior of individuals is probabilistic. Each agent independently chooses the first behavior with probability $\bar{p}$ (the mean of the Bernoulli distribution) and  the alternative behavior with complementary probability $1-\bar{p}$ at each time step.

The literature shows that the impact of the dynamics choice is highly model-dependent.
For example, it changes the bifurcation types or modifies the system sensitivity to external parameters in the the nonlinear $q$-voter model \cite{Jed:Szn:20}, certain variants of the $q$-voter model with independence \cite{Szn:Szw:Wer:14, Jed:Szn:17, Byr:etal:16}, and decision-making models with symmetric social learning functions \cite{Jed:Her:24}.
In contrast, other models are entirely unaffected by the choice of the dynamics, like some variants of the $q$-voter model with anticonformity \cite{Jed:Szn:17,Pra:Mul:Sen:26} and models of decision-making with nonsymmetric conformity functions \cite{Jed:Her:24}, or are affected only quantitatively, yielding
qualitatively similar results, as in the Galam model with contrarians \cite{Sta:Mar:04} and the Sznajd model \cite{Sch:04}. 

\section{Analysis of the Framework}
\label{sec:anal}
As reviewed in Section~\ref{sec:equi}, the equivalence between annealed and quenched dynamics is highly model-dependent.
A heterogeneous system under quenched dynamics cannot always be reduced to a homogeneous system characterized only by a mean preference as its behavior often depends on the entire shape of its preference distribution.
Consequently, two populations that share the exact same mean preference but are modeled by different underlying distributions can evolve towards completely different stationary states.
Thus, no single mean value is sufficient to uniquely characterize such a heterogeneous population.
However, there is a special condition under which this is possible \cite{Jed:Men:25}.

To derive this condition, we track the fraction of agents in state $A$, denoted by $a$.
By definition, the remaining fraction of the population in state $B$ is $1-a$.
We derive the differential equations that describe how $a$ changes over time and determine the fixed points of the dynamics, which corresponds to the final states of the system.
Sections~\ref{sec:annealed} and \ref{sec:quenched} present the analysis for the annealed and quenched dynamics, respectively, whereas Section~\ref{sec:bal} reveals the balancing condition under which both the dynamics become equivalent.

\subsection{Annealed Dynamics}
\label{sec:annealed}
Annealed dynamics means that the preference $p_i$ of each agent is redrawn independently from the preference distribution, $\phi(p)$, at every time step, see Fig.~\ref{fig:que-ann-vis:a}.
The fraction of agents in state $A$ changes over time, and the rate of this change is given by
\begin{equation}
\label{eq:rate-equation}
    \frac{da}{dt}=(1-a)P_{BA}-aP_{AB},
\end{equation}
where the first and second terms represent the average macroscopic inflow into and outflow from state $A$, respectively. 

For agent $i$ with a given preference $p_i$, the probabilities of switching states from $B$ to $A$ and from $A$ to $B$ are defined by
\begin{equation}
\begin{split}
    P_{BA}(p_i)&=p_i X_{BA}(a)+(1-p_i)Y_{BA}(a),\\
    P_{AB}(p_i)&= p_i X_{AB}(a)+(1-p_i)Y_{AB}(a),
\end{split}
\end{equation}
which follows from the construction of the modeling framework, see Section~\ref{sec:model}.
Since the preferences of all agents change independently at every time step, the macroscopic transition probabilities from Eq.~\eqref{eq:rate-equation} correspond to the expected values of the above individual probabilities over the preference distribution $\phi(p)$:
\begin{equation}
\begin{split}
    P_{BA}=&\int P_{BA}(p)\phi(p)dp,\\
    P_{AB}=&\int P_{AB}(p)\phi(p)dp.
\end{split}  
\end{equation}
Because $P_{AB}(p)$ and $P_{BA}(p)$ are linear in $p$, we eventually get:
\begin{equation}
\label{eq:transition-rates-2}
\begin{split}
    P_{BA}&=\bar{p} X_{BA}(a)+(1-\bar{p})Y_{BA}(a),\\
    P_{AB}&= \bar{p} X_{AB}(a)+(1-\bar{p})Y_{AB}(a),
\end{split}
\end{equation}
where $\bar{p}$ is the mean preference, i.e., $\bar{p}=\int p \phi(p)dp$.
Thus, under annealed dynamics, the collective behavior depends only on the mean of the preference distribution, leaving its overall shape irrelevant.
A heterogeneous population behaves identically to a homogeneous one where all agents share the same preference $\bar{p}$.

Substituting Eqs.~\eqref{eq:transition-rates-2} into Eq.~(\ref{eq:rate-equation}), we get the time evolution of $a$:
\begin{equation}
\label{eq:time_evol_ann}
\begin{split}
    \frac{da}{dt}=&\bar{p}X_{BA}(a)+(1-\bar{p})Y_{BA}(a)-\left[Y_{BA}(a)+Y_{AB}(a)\right]a\\
    &-\left[X_{BA}(a)-Y_{BA}(a)+X_{AB}(a)-Y_{AB}(a)\right]\bar{p}a.
\end{split}
\end{equation}
Since Eq.~\eqref{eq:time_evol_ann} describes a one-dimensional autonomous flow, the system cannot exhibit oscillatory behavior \cite{Str:18}.
Any trajectory $a(t)$ is monotonic and asymptotically approaches a stationary state.

The stationary states are found by identifying fixed points, which are denoted by $a^*$ and defined by the condition:
\begin{equation}
\label{eq:fixed-points-annealed}
    \left.\frac{da}{dt}\right\vert_{a^*}=0.
\end{equation}
Combining Eqs.~\eqref{eq:time_evol_ann} and \eqref{eq:fixed-points-annealed}, the fixed points are given by the following implicit formula:
\begin{equation}
\label{eq:fixed-points-ann}
    \bar{p}=\frac{Y_{BA}(a^*)-a^*\left[Y_{AB}(a^*)+Y_{BA}(a^*)\right]}{Y_{BA}(a^*)-X_{BA}(a^*)+a^*\left[X_{AB}(a^*)+X_{BA}(a^*)-Y_{AB}(a^*)-Y_{BA}(a^*)\right]}.
\end{equation}

\subsection{Quenched Dynamics}
\label{sec:quenched}
Quenched dynamics means that each agent is assigned an individual preference, $p_i$, drawn from the distribution $\phi(p)$, at $t=0$, and this preference remains fixed throughout the entire time evolution, see Fig.~\ref{fig:que-ann-vis:b}.
To describe such a system, we divide the population into groups of agents sharing the same preference, and we write down a septate rate equation for each of them.

Let $a_p$ denote the fraction of agents in state $A$ within the group characterized by preference $p$.
Equivalently, $a_p$ represents the conditional probability that an agent with preference $p_i=p$ is in state $A$.
By the law of total probability, the global fraction of agents in state $A$ in the entire population is the expected value of $a_p$ over the preference distribution:
\begin{equation}
\label{eq:apopulation}
    a=\int a_p\phi(p)dp.
\end{equation}

In analogy to Eq.~\eqref{eq:rate-equation}, the rate equation for the group characterized by preference $p$ is given by:
\begin{equation}
    \label{eq:rate-que}
    \frac{da_p}{dt}=(1-a_p)P^p_{BA}-a_pP^p_{AB},
\end{equation}
where $P^p_{BA}$ and $P^p_{AB}$ are the transition probabilities with the following forms:
\begin{equation}
\begin{split}
\label{eq:transition-rates-que}
    P^p_{BA}&=pX_{BA}(a)+(1-p)Y_{BA}(a),\\
    P^p_{AB}&= pX_{AB}(a)+(1-p)Y_{AB}(a).
\end{split}  
\end{equation}
While the annealed dynamics is strictly constrained to a one-dimensional flow, Eq.~\eqref{eq:time_evol_ann}, the quenched dynamics of a heterogeneous system generally describes a high-dimensional flow governed by the set of coupled differential equations, Eq.~\eqref{eq:rate-que}.
In such a system, an oscillatory behavior may arise \cite{Str:18}.
This stands in contrast to homogeneous systems or heterogeneous systems under annealed dynamics, where oscillations are not possible.

To derive the equation for the time evolution of $a$, we differentiate Eq.~\eqref{eq:apopulation} with respect to time. 
Since the preference distribution is time-independent, the derivative commutes with the integral yielding
\begin{equation}
\label{eq:rate-que-t}
    \frac{da}{dt}=\int\frac{da_p}{dt}\phi(p)dp.
\end{equation}
Combining Eqs.~\eqref{eq:rate-que}-\eqref{eq:rate-que-t}, we get:
\begin{equation}
\label{eq:dadt_que}
\begin{split}
    \frac{da}{dt}=&\bar{p}X_{BA}(a)+(1-\bar{p})Y_{BA}(a)-\left[Y_{BA}(a)+Y_{AB}(a)\right]a\\
    &-\left[X_{BA}(a)-Y_{BA}(a)+X_{AB}(a)-Y_{AB}(a)\right]\int pa_p\phi(p)dp.
\end{split}
\end{equation}
To perform the integral in the last term of Eq.~\eqref{eq:dadt_que}, we need to know the entire preference distribution, $\phi(p)$.
This is in contrast to the annealed case, where the dynamics is fully determined only by the mean preference, $\bar{p}$, see Eq.~\eqref{eq:time_evol_ann}.

The stationary states of the system are found by identifying fixed points $\{a_p^*\}$, for which
\begin{equation}
    \label{eq:dap}
    \left.\frac{da_p}{dt}\right\vert_{\{a_p^*\}}=0.
\end{equation}
From Eqs.~(\ref{eq:rate-que}), (\ref{eq:transition-rates-que}), and (\ref{eq:dap}), the fixed points are determined by:

\begin{equation}
\label{eq:ap_que}
    a_p^*=\frac{Y_{BA}(a^*)-p\left[Y_{BA}(a^*)-X_{BA}(a^*)\right]}{Y_{BA}(a^*)+Y_{AB}(a^*)+p\left[X_{BA}(a^*)+X_{AB}(a^*)-Y_{BA}(a^*)-Y_{AB}(a^*)\right]}.
\end{equation}
Having Eq.~\eqref{eq:ap_que}, the stationary fraction $a^*$ of agents in state $A$ is obtained by solving the self-consistency equation:
\begin{equation}
\label{eq:apopulation-st}
    a^*=\int a_p^*\phi(p)dp.
\end{equation}
In general, this equation cannot be integrated to solve for $a^*$ without specifying the exact form of the preference distribution, $\phi(p)$.
In \ref{sec:app}, we present the calculations for  degenerate and Bernoulli distributions.

\subsection{Balancing Condition}
Although the quenched system cannot be fully described by the mean preference and mapped into a homogeneous system in general, it is possible in special cases.
\label{sec:bal}
When the transition probabilities defining the model, see Section~\ref{sec:model}, satisfy the balancing condition given by
\begin{equation}
\label{eq:balance}
     X_{AB}(a)+X_{BA}(a)=Y_{AB}(a)+Y_{BA}(a)
\end{equation}
for all $a\in[0,1]$, the quenched dynamics becomes equivalent to the annealed one in heterogeneous systems \cite{Jed:Men:25}.
This is evident as under Eq.~\eqref{eq:balance}, the differential equations for annealed and quenched dynamics, Eqs.~\eqref{eq:time_evol_ann} and \eqref{eq:rate-que}, simplify to
\begin{equation}
\label{eq:da-balance}
\begin{split}
    \frac{da}{dt}=\bar{p}X_{BA}(a)+(1-\bar{p})Y_{BA}(a)-\left[Y_{BA}(a)+Y_{AB}(a)\right]a.
\end{split}
\end{equation}
Because Eq.~\eqref{eq:da-balance} depends exclusively on the mean preference $\bar{p}$, the full shape of the preference distribution becomes irrelevant.
This allows any heterogeneous population to be mapped directly into an equivalent homogeneous system, see Fig.~\ref{fig:diagram}.

Setting Eq.~\eqref{eq:da-balance} to zero gives the implicit formula for the fixed-points:
\begin{equation}
\label{eq:fixed-points-balance}
    \bar{p}=\frac{Y_{BA}(a^*)-a^*\left[Y_{BA}(a^*)+Y_{AB}(a^*)\right]}{Y_{BA}(a^*)-X_{BA}(a^*)},
\end{equation}
which can also be derived by directly applying the balancing condition to Eq.~\eqref{eq:fixed-points-ann} in the case of the annealed dynamics or to Eqs.~\eqref{eq:ap_que} and \eqref{eq:apopulation-st} in the case of the quenched dynamics.

Thus, under the balancing condition, the annealed and quenched dynamics share identical time evolution and stationary states.
This equivalence rules out the possibility of oscillations in heterogeneous systems that satisfy the balancing condition.

\section{Applications of the Framework}
\label{sec:dis}
The presented framework covers various models from different disciplines.
In this section, we present some examples of how it can be used to better understand the impact of heterogeneity as well as the annealed and quenched dynamics on the model behavior.
Section~\ref{sec:socio} focuses on models of social influence, Section~\ref{sec:spin} on spin models, whereas Section~\ref{sec:dise} on models of disease spreading.

\subsection{Models of Social Influence}
\label{sec:socio}
In models of social system, mechanisms $X$ and $Y$ may represent different types of social response \cite{Jed:Szn:19}, heuristic rules \cite{Oli:92,Vie:etal:20}, or learning strategies \cite{Yan:etal:21,Wu:etal:25} that dictate agent behavior.
Let us consider a nonlinear voter model with anticonformity defined by the following functions: 
\begin{equation}
    \begin{aligned}
        X_{BA}(a)=&(1-a)^{\alpha}, \\
        X_{AB}(a)=&a^{\alpha}, \\
        Y_{BA}(a)=&ra^{\beta}, \\
        Y_{AB}(a)=&r(1-a)^{\beta}.
    \end{aligned}
    \begin{aligned}
    &\left.\vphantom{\begin{aligned}
        X_{BA}(a)=&(1-a)^{\alpha}, \\
        X_{AB}(a)=&a^{\alpha}, \\
      \end{aligned}}\right\rbrace\quad\text{Anticonformity}\\
    &\left.\vphantom{\begin{aligned}
        Y_{BA}(a)=&a^{\beta}, \\
        Y_{AB}(a)=&(1-a)^{\beta}.
      \end{aligned}}\right\rbrace\quad\text{Conformity}
    \end{aligned}
    \end{equation}
The mechanisms $X$ and $Y$ represent anticonformity and conformity, respectively.
The exponents $\alpha$ and $\beta$ control the nonlinearity of the corresponding social responses.
The parameter $r$ scales the intensity of conformity relative to anticonformity: both the mechanisms have the same intrinsic strength for $r=1$, $r>1$ strengthens the impact of conformity, whereas $r<1$ weakens it.
Finally, $p_i$ reflects the agents' preference towards anticonfomrity. 
Depending on the choice of parameters, this formulation reproduces several nonlinear voter models studied previously \cite{Nyc:Szn:Cis:12,Nyc:Szn:13,Jed:Szn:17,Jed:Szn:22,Mas:13,Tan:Mas:13,Abr:Paw:Szn:19,Abr:etal:21}.

We consider four preference distributions with the same mean $\bar{p}=0.5$: a degenerate distribution, a Bernoulli distribution, a uniform distribution on $[0,1]$, and a truncated normal distribution with variance $\sigma^2=1/40$.
Figure~\ref{fig:distributions} illustrates these distributions.
The degenerate distribution corresponds to a homogeneous population, and thus it leads to identical system behavior under the annealed and quenched dynamics. 
It serves as a reference for interpreting later results.
The rest three distributions model a heterogeneous population, where agents differ in their preferences.
The Bernoulli distribution is a special one among them as it effectively divides a population into two groups characterized by different behaviors.
One only follows the mechanisms $X$ (anticonformity), whereas the other only follows the mechanisms $Y$ (conformity).

\begin{figure}[!t]
	\subfloat{\label{fig:tra:a}}
	\subfloat{\label{fig:tra:b}}
	\subfloat{\label{fig:tra:c}}
	\subfloat{\label{fig:tra:d}}
	\subfloat{\label{fig:tra:e}}
	\subfloat{\label{fig:tra:f}}
	\centering
	\includegraphics[width=\linewidth]{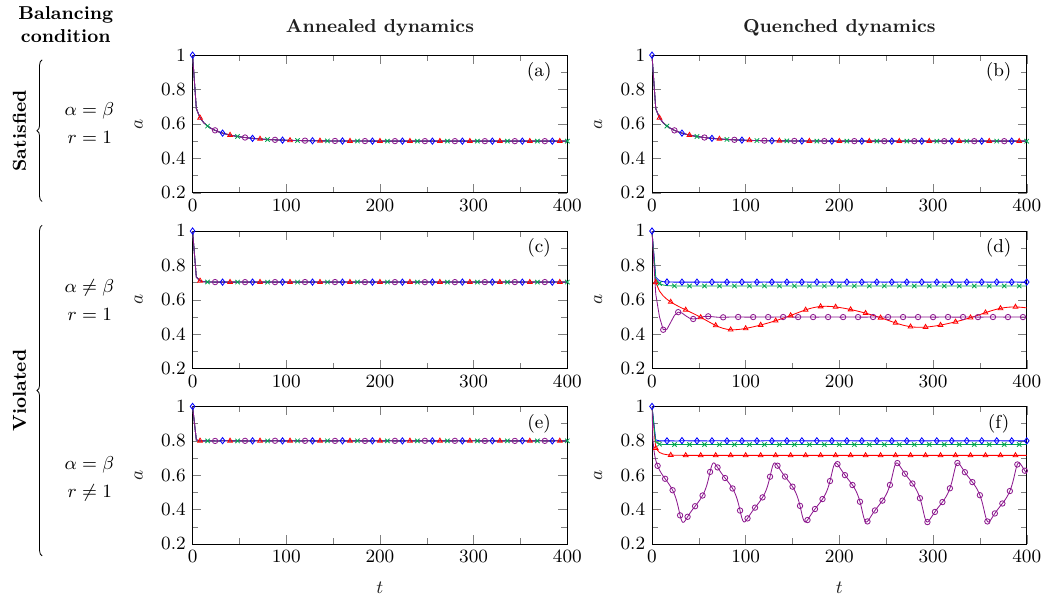}
	\caption{\label{fig:trajectories} Time evolution of different variants of the nonlinear voter model with anticonfomrity for four preference distributions with the same mean $\bar{p}=0.5$ that are presented in Fig.~\ref{fig:distributions}.
		Plots (a) and (b) illustrate the behavior of models that satisfy the balancing condition ($\alpha=\beta=2$, $r=1$). The rest correspond to models that violate it: (c) and (d) $\alpha=6$, $\beta=2$, $r=1$, and (d) and (e) $\alpha=6$, $\beta=2$, $r=1$.
		The distributions are represented by unique colors and symbols from Fig.~\ref{fig:distributions}.
	}
\end{figure}

When $\alpha=\beta$ and $r=1$ \cite{Nyc:Szn:Cis:12,Nyc:Szn:13,Jed:Szn:17,Jed:Szn:22,Tan:Mas:13} (Figs.~\ref{fig:tra:a} and \ref{fig:tra:b}), the balancing condition is satisfied. 
The annealed and quenched dynamics are equivalent, and the time evolution of the system is determined entirely by the mean preference $\bar{p}$.
Consequently, all four distributions lead to identical trajectories for both the dynamics.
This means that any heterogeneous population can be effectively mapped into a homogeneous one with the corresponding mean preference.
Moreover, such a system cannot exhibit any oscillatory behavior.

When $\alpha\neq\beta$ and $r=1$ \cite{Abr:Paw:Szn:19,Abr:etal:21} (Figs.~\ref{fig:tra:c} and \ref{fig:tra:d}) or $\alpha=\beta$ and $r\neq1$ \cite{Tan:Mas:13} (Figs.~\ref{fig:tra:e} and \ref{fig:tra:f}), the balancing condition is violated.
The annealed dynamics still depends only on the mean preference, so the trajectories for different distributions coincide in Figs.~\ref{fig:tra:c} and \ref{fig:tra:e}.
However, the mean preference is not enough to describe the quenched dynamics, and we see that the shape of the distribution drastically impacts the trajectories of the system in Figs.~\ref{fig:tra:d} and \ref{fig:tra:f}.
Depending on a distribution, the system may converge to different fixed points in a monotonic or oscillatory way, or even exhibit sustained oscillations.
In this case, annealed and quenched dynamics are not equivalent, and we cannot map a heterogeneous population into a homogeneous one.
Moreover, quenched dynamics may exhibit oscillations.

The behavior described above helps us reinterpret some earlier findings on specific formulations of this model.
In particular, the version that satisfies the balancing condition ($\alpha=\beta$ and $r=1$) was reported to give the same results in a setting of a homogeneous population (annealed case) and a heterogeneous one with Bernoulli-distributed preferences (quenched case) in Ref.~\cite{Jed:Szn:17}.
Our analysis provides a theoretical explanation for these observations.
The equivalence of annealed and quenched dynamics for the nonlinear voter model with anticonformity follows directly from the balancing condition, which ensures that the effects of preference heterogeneity cancel out at the population level.
Therefore, the system is invariant under any form of heterogeneity, not only under the one introduced by the Bernoulli distribution considered in  Ref.~\cite{Jed:Szn:17}. 
However, it is important to note that anticonformity, as well as other social responses, can take various mathematical formulations \cite{Cla:etal:14,Cla:Whi:12,Cla:Bow:Whi:12,Mor:Lal:12}, and different representations can either satisfy or violate the balancing condition, as seen in the given example.
Therefore, robustness to heterogeneity should not be regarded as a universal property of anticonformity itself but rather as a feature of specific model parametrization.

This observation underlines that in agent-based modeling, the implication of socio-psychological mechanisms depend crucially on how they are operationalized mathematically.
Similarly, completely different social mechanisms can lead to identical update formulas, so one must be cautious when using macroscopic patterns to justify or validate chosen microscopic rules \cite{Gal:05,Gal:22}. 

\subsection{Spin Models with Competing Dynamics}
\label{sec:spin}
The investigation of kinetic Ising models with competing dynamics has been instrumental in characterizing non-equilibrium phase transitions.
These models typically feature either a competition between distinct dynamics---like a single-flip Glauber and a spin-exchange Kawasaki dynamics \cite{Gon:Gar:Leb:87,Tom:Oli:89,Dum:God:24}---or a competition within a single dynamics operating under locally competing temperatures \cite{Gar:Lab:Mar:87, Tom:Oli:San:91, Tam:Ale:Gup:94}.
Such systems have been studied primary under annealed dynamics, leaving the impact of population heterogeneity largely unexplored.

In models with competing temperatures, the system is in contact with two thermal baths simultaneously. 
We consider a kinetic Ising model where the competition is defined by the following: with probability $p$, a spin interacts with $q_1$ randomly chosen neighbors at temperature $T_1$, whereas with probability $1-p$, it interacts with $q_2$ randomly chosen neighbors at temperature $T_2$.
For the Metropolis dynamics, the model is defined by:
\begin{equation}
            \begin{aligned}
                X_{BA}(a)=&\sum_{i=0}^{q_1}\binom{q_1}{i}a^i(1-a)^{q_1-i}\min\left[1,e^{2\beta_1(2i-q_1)}\right], \\
                X_{AB}(a)=&\sum_{i=0}^{q_1}\binom{q_1}{i}a^i(1-a)^{q_1-i}\min\left[1,e^{-2\beta_1(2i-q_1)}\right], \\
                Y_{BA}(a)=&\sum_{i=0}^{q_2}\binom{q_2}{i}a^i(1-a)^{q_2-i}\min\left[1,e^{2\beta_2(2i-q_2)}\right], \\
                Y_{AB}(a)=&\sum_{i=0}^{q_2}\binom{q_2}{i}a^i(1-a)^{q_2-i}\min\left[1,e^{-2\beta_2(2i-q_2)}\right],
            \end{aligned}
            \begin{aligned}
            &\left.\vphantom{\begin{aligned}
                X_{BA}(a)=&\sum_{i=0}^{\frac{n-1}{2}}\binom{n}{i}a^i(1-a)^{n-i}, \\
                X_{AB}(a)=&\sum_{i=\frac{n+1}{2}}^{n}\binom{n}{i}a^i(1-a)^{n-i}, \\
              \end{aligned}}\right\rbrace\quad\text{\text{Heat bath at $T_1$}}\\
            &\left.\vphantom{\begin{aligned}
                Y_{BA}(a)=&\sum_{i=\frac{n+1}{2}}^{n}\binom{n}{i}a^i(1-a)^{n-i}, \\
                Y_{AB}(a)=&\sum_{i=0}^{\frac{n-1}{2}}\binom{n}{i}a^i(1-a)^{n-i}.
              \end{aligned}}\right\rbrace\quad\text{\text{Heat bath at $T_2$}}
            \end{aligned}
\end{equation} 
and it does not satisfy the balancing condition.
In contrast, the analogous model with Glauber dynamics is defined as follows:
\begin{equation}
            \begin{aligned}
                X_{BA}(a)=&\frac{1}{2}\sum_{i=0}^{q_1}\binom{q_1}{i}a^i(1-a)^{q_1-i}\left\{1+\tanh\left[\beta_1(2i-q_1)\right]\right\}, \\
                X_{AB}(a)=&\frac{1}{2}\sum_{i=0}^{q_1}\binom{q_1}{i}a^i(1-a)^{q_1-i}\left\{1-\tanh\left[\beta_1(2i-q_1)\right]\right\}, \\
                Y_{BA}(a)=&\frac{1}{2}\sum_{i=0}^{q_2}\binom{q_2}{i}a^i(1-a)^{q_2-i}\left\{1+\tanh\left[\beta_2(2i-q_2)\right]\right\}, \\
                Y_{AB}(a)=&\frac{1}{2}\sum_{i=0}^{q_2}\binom{q_2}{i}a^i(1-a)^{q_2-i}\left\{1-\tanh\left[\beta_2(2i-q_2)\right]\right\},
            \end{aligned}
            \begin{aligned}
            &\left.\vphantom{\begin{aligned}
               X_{BA}(a)=&\frac{1}{2}\sum_{i=0}^{q_1}\binom{q_1}{i}a^i(1-a)^{q_1-i}\left\{1+\tanh\left[\beta_1(2i-q_1)\right]\right\}, \\
                X_{AB}(a)=&\frac{1}{2}\sum_{i=0}^{q_1}\binom{q_1}{i}a^i(1-a)^{q_1-i}\left\{1-\tanh\left[\beta_1(2i-q_1)\right]\right\}, \\
              \end{aligned}}\right\rbrace\quad\text{\text{Heat bath at $T_1$}}\\
            &\left.\vphantom{\begin{aligned}
                Y_{BA}(a)=&\frac{1}{2}\sum_{i=0}^{q_2}\binom{q_2}{i}a^i(1-a)^{q_2-i}\left\{1+\tanh\left[\beta_2(2i-q_2)\right]\right\}, \\
                Y_{AB}(a)=&\frac{1}{2}\sum_{i=0}^{q_2}\binom{q_2}{i}a^i(1-a)^{q_2-i}\left\{1-\tanh\left[\beta_2(2i-q_2)\right]\right\}.
              \end{aligned}}\right\rbrace\quad\text{\text{Heat bath at $T_2$}}
            \end{aligned}
\end{equation} 
and it satisfies the balancing condition.

For the special case where $q_1=q_2$ and $T_1=T_2$, we recover the $q$-neighbor Ising model \cite{Jed:Chm:Szn:15, Par:Noh:17,Jed:Chm:Szn:17, Chm:Gra:Kra:18,Kra:21}.
Under Metropolis dynamics, this baseline model exhibits a notable behavior associated with the hysteresis. For $q=3$, the phase transition is continuous. 
However, for $q\geq4$, it becomes discontinuous with oscillating width of the hysteresis, expanding for even values of $q$ and shrinking for odd ones \cite{Jed:Chm:Szn:15}.
On the other hand, the baseline model with Glauber dynamics exhibits only continuous phase transitions \cite{Jed:Chm:Szn:17}.

\begin{figure}[!t]
	\subfloat{\label{fig:ising:a}}
	\subfloat{\label{fig:ising:b}}
	\subfloat{\label{fig:ising:c}}
	\subfloat{\label{fig:ising:d}}
	\subfloat{\label{fig:ising:e}}
	\subfloat{\label{fig:ising:f}}
	\centering
	\includegraphics[width=\linewidth]{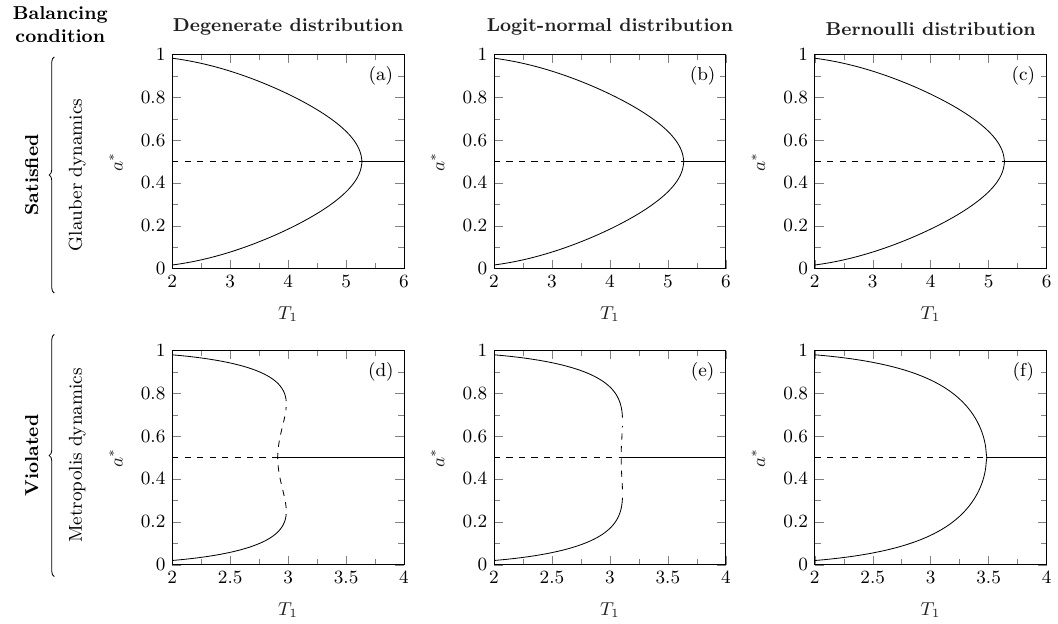}
	\caption{\label{fig:ising} Phase diagrams for the model with $q_1=4$, $q_2=3$, $T_2=1$ under the quenched dynamics. Each column corresponds to a different distribution of $p$ with the same mean $\bar{p}=0.5$, from left to right: degenerate distribution, logit-normal distribution (with location and squared scale parameters: $\mu=0$ and $\sigma^2=4$), and Bernoulli distribution. 
		Note that for the degenerate distribution, annealed and quenched dynamics are equivalent. 
		The top row illustrates the behavior of the models with competing Glauber dynamics (balancing condition satisfied). In contrast, the bottom row illustrates the behavior of the models with competing Metropolis dynamics (balancing condition violated). 
	}
\end{figure}

Figure~\ref{fig:ising} illustrates the behavior of both the models with $q_1=4$ and $q_2=3$, where the temperature of the first heat bath, $T_1$, is varied, while the temperature of the second is fixed at $T_2=1$.  
We consider three distributions of the parameter $p$: a degenerate distribution, a log-normal distribution, and a Bernoulli distribution. 
Only the last two distributions model a heterogeneous system.
Although these diagrams present the system under quenched dynamics, the first column (with the degenerate distribution) simultaneously represents the annealed dynamics, as the annealed and quenched dynamics are equivalent in homogeneous systems.

Because the Glauber variant satisfies the balancing condition, its phase diagram remain invariant across all there distributions, see Figs.~\ref{fig:ising:a}-\ref{fig:ising:c}.
In contrast, the Metropolis variant, which violates the balancing condition, exhibits strong sensitivity to the underlying distribution, to the extent that the nature of the phase transition changes. 
Specifically, while the discontinuous phase transition is merely softened under the log-normal distribution compared to the degenerate case, see Figs.~\ref{fig:ising:d} and \ref{fig:ising:e}, it is entirely eliminated under the Bernoulli distribution, where the transition becomes continuous, see Fig.\ref{fig:ising:f}.

Such softening of a discontinuous phase transition or its transformation into a continuous one under the quenched dynamics is known as a rounding effect \cite{Aiz:Weh:89,Bor:Mar:Mun:13,Vil:Bon:Mun:14}.
Our results show that the specific profile of the disorder distribution directly dictates whether the discontinuity is simply weakened or completely destroyed.

\subsection{Models of Disease Spreading}
\label{sec:dise}
Our framework covers also models of disease spreading. Let $a$ denote the fraction of infected agents, whereas $1-a$ the fraction of susceptible ones. Then, the Susceptible-Infected-Susceptible (SIS) model \cite{Kee:Eam:05,Bra:08,Bar:Bar:Ves:08}, can be defined as
\begin{equation}
    \begin{aligned}
        X_{BA}(a)=&\beta a, \\
        X_{AB}(a)=&0, \\
        Y_{BA}(a)=&0, \\
        Y_{AB}(a)=&\gamma.
    \end{aligned}
    \begin{aligned}
    &\left.\vphantom{\begin{aligned}
        X_{BA}(a)=&\beta(1-a), \\
        X_{AB}(a)=&0, \\
      \end{aligned}}\right\rbrace\quad\text{Infection}\\
    &\left.\vphantom{\begin{aligned}
        Y_{BA}(a)=&0, \\
        Y_{AB}(a)=&\gamma.
\end{aligned}}\right\rbrace\quad\text{Recovery}
    \end{aligned}
    \end{equation}
The mechanisms $X$ and $Y$ represent infection and recovery, respectively.
The disease can be transmitted from an infected person to a susceptible one when they come into contact. The parameter $\beta$ controls this transmission rate. 
Infected individuals recover at  a constant rate controlled by $\gamma$ regardless of their contacts with other agents.

Under the annealed dynamics, this formulation leads to the differential equation known for SIS model \cite{Kee:Eam:05}. Using Eq.~\eqref{eq:time_evol_ann}, the time evolution of the infected fraction is given by
\begin{equation}
\label{eq:time_sis}
    \frac{da}{dt}=\beta \bar{p}a(1-a)-\gamma(1-\bar{p})a,
\end{equation}
which can be rewritten in the conventional epidemiological form
\begin{equation}
    \frac{dI}{dt}=\beta'IS-\gamma'I,
\end{equation}
where  $I=a$, $S=1-a$, and the effective transmission and recovery rates are defined as $\beta'=\beta \bar{p}$ and $\gamma'=\gamma(1-\bar{p})$, respectively.
The ratio of these effective rates defines the basic reproductive number \cite{Bra:08,Bar:Bar:Ves:08}, 
\begin{equation}
    R_0=\frac{\beta'}{\gamma'}=\frac{\beta\bar{p}}{\gamma(1-\bar{p})},
\end{equation}
which serves as a critical threshold that determines the stability of the disease-free equilibrium ($a=0$). When $R_0<1$, the recovery process dominates, and the infection dies out.
In contrast, for $R_0>1$, the disease persists in a stationary fraction $a^*=1-1/R_0$. The parameter $\bar{p}$ shifts the epidemic threshold based on the relative frequency of infection and recovery.

\begin{figure}[!t]
	\subfloat{\label{fig:epi:a}}
	\subfloat{\label{fig:epi:b}}
	\subfloat{\label{fig:epi:c}}
	\centering
	\includegraphics[width=\linewidth]{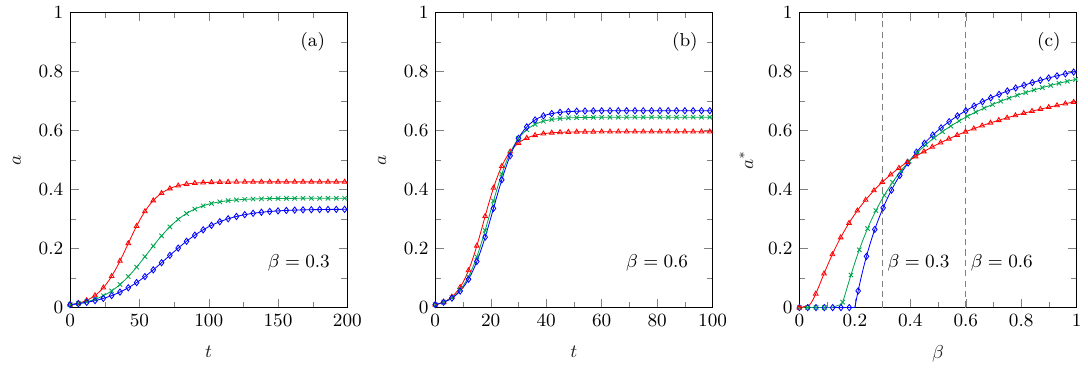}
	\caption{\label{fig:epidemics} Time evolution (a,b) and stable fixed points (c) for the SIS model with $\gamma=0.2$ for three preference distributions with the same mean $\bar{p}=0.5$: one-point distribution (blue $\diamond$), uniform distribution on $[0,1]$ (red $\triangle$), and normal distribution centered at $p=0.5$ with variance $\sigma^2=1/40$ (green $\times$), which are also presented in Fig.~\ref{fig:distributions} with the same colors and symbols.
	}
\end{figure}

The SIS model does not satisfy the balancing condition. 
Thus, introducing heterogeneity by assigning different values of $p$ to agents will alter the dynamics of disease spread.
Figure~\ref{fig:epidemics} illustrates this by presenting the behavior of the model for three different frequency distributions with the same mean $\bar{p}=0.5$: a one-point distribution, a uniform distribution, and a normal distributions. The distributions used are presented in Fig.~\ref{fig:distributions}.

Figures~\ref{fig:epi:a} and \ref{fig:epi:b} depict the time evolution of the infected fraction $a$ for two transmission rates, $\beta=0.3$ and $\beta=0.6$, respectively. 
For lower transmission rates ($\beta=0.3$), the uniform distribution, characterized by the highest variance, leads to the fastest increase in infections and the highest steady-state level. 
In contrast, the one-point distribution, representing a homogeneous population, results in the lowest level of infection.
However, for higher transmission rates ($\beta=0.6$), we observe the opposite behavior.
The homogeneous population eventually reaches a higher level of infection than the heterogeneous ones.

Thus, while population diversity can accelerate the initial disease spread, it can also provide a form of resistance against a high infection level when the pathogen is highly contagious.
This can be further seen in Fig.~\ref{fig:epi:c}, which shows the stable infection levels, $a^*$, as a function of the transmission rate $\beta$. 
Heterogeneity lowers the epidemic threshold. Nevertheless, it also reduces the final epidemic size in highly transmissive scenarios.

While the above analysis focuses on the SIS model, our formalism can describe a border class of epidemic processes. 
For example, by setting $\gamma=0$, there is no possibility of recovery, so we get the Susceptible-Infected (SI) model \cite{Bra:08,Bar:Bar:Ves:08}.
Furthermore, the discussed example can be extended to social contagion models \cite{Hil:etal:10} by introducing a spontaneous infection rate $\alpha$, such that $X_{BA}(a)=\beta a +\alpha$ and $Y_{BA}(a)=\alpha$.

\section{Limitations of the Framework}
\label{sec:limit}
The presented formalism relies on a few simplifying assumptions.
This section highlights the limitations of the framework and outlines corresponding avenues of its future development.

\subsection{Networked Populations}
The presented balancing condition is derived for well-mixed populations. 
Introducing the network topology may significantly alter its form. 
Actually, previous studies indicate that the differences between quenched and annealed dynamics appear on networks even though a model satisfies the balancing condition in a well-mixed population \cite{Jed:Szn:22}.
These differences are particularly pronounced in networks with a small average node degree, and as the average degree increases, the results converge towards a well-mixed population. 

These results suggest that the implications of satisfying the balancing condition may still apply to sufficiently dense networks.
However, the specific role of the network structure is likely model-dependent, so more studies are needed in this direction.
To analytically capture these effects, approximations developed for network problems \cite{Gle:13}, like the pair approximation for binary-choice  \cite{Jed:17,Jed:Szn:22} and multi-state dynamics \cite{Lip:Szn:25}, look promising.

\subsection{Multi-state Models}
Our framework is restricted to binary-choice models.
Although these models are widely studied, many of them have been extended to accommodate more than two states \cite{Ram:etal:24,Lip:Szn:22,Don:Lip:Szn:22,Now:Szn:22,Lip:Szn:25,Now:Sto:Szn:21,Szn:Szn:Wer:21}.
Expanding the state space is particularly useful in opinion dynamics to study polarization, as the multi-state format of opinion naturally reflects a discrete Likert scale frequently employed in empirical research \cite{Lip:Szn:22,Don:Lip:Szn:22}. 
However, the equivalence of annealed and quenched dynamics guaranteed by the balancing condition does not necessary translate to these multi-state extensions.

For example, while the binary-choice $q$-voter model with anticoformity fulfills the balancing condition, its multi-state counterpart is sensitive to the choice of the dynamics \cite{Now:Szn:22,Lip:Szn:25}.
Moreover, the change of the dynamics from annealed to quenched in certain multi-sate models may induce a discontinuous phase transition where a continuous one previously existed \cite{Now:Szn:22}.
This behavior directly opposes the rounding effect \cite{Aiz:Weh:89,Bor:Mar:Mun:13,Vil:Bon:Mun:14}, showing that multi-state systems can reveal novel phenomena.

\subsection{Multi-flip Dynamics}
The current framework assumes a random sequential updating scheme, where only a single agent changes its state at a time.
However, some models incorporate multi-agent updates. 
For example, the Sznajd model involves simultaneous updates by a pair of agents \cite{Szn:Szn:Wer:21}, while group updates are central to the Galam model \cite{Gal:26, Gal:04, Sta:Mar:04}.
Furthermore, our framework does not account for a synchronous (parallel) updating scheme, where the entire population updates their states simultaneously \cite{Gra:Li:20,Juu:Por:19,Jav:Squ:15}.

\subsection{Multiple Competing Mechanisms and Heterogeneous Parameters}
Our framework covers only competitions between exactly two mechanisms.
However, systems may be driven by an interplay between more distinct mechanisms.
Examples include the simultaneous integration of conformity, anticonformity, and independence in the $q$-voter model \cite{Nyc:etal:18} or the majority-vote model \cite{Vie:Cro:16,Jes:etal:19,Kra:Gra:19} as well as combinations involving  media influence \cite{Civ:21} or skepticism \cite{Anu:etal:25}.

Additionally, this work focuses on heterogeneity related to the mechanism selection probability.
However, other model parameters may be considered to be quenched or annealed.
For example, thresholds in the threshold model are introduced as quenched properties \cite{Now:Gra:Szn:22,Gra:Li:20}.
    
\section{Conclusions}
\label{sec:con}
A fundamental pursuit in physics is the search for unifying frameworks and universal theories capable of explaining different phenomena through a single set of core principles.
The proposed framework \cite{Jed:Men:25} contributes to both of these goals.
First, it exposes structural similarities between models that originate in different disciplines.
Second, it reveals a simple law---the balancing condition on tradition rates---which dictates whether a system is sensitive to the details of its dynamics and heterogeneity, thereby resolving a long-standing puzzle in the literature.

In this article, we analyze the framework in detail and present its cross-disciplinary applications.
The biggest strength of the framework is its ability to compare directly different approaches commonly used in modeling various complex systems: annealed and quenched dynamics together with homogeneous and heterogeneous populations of agents.
Our literate review reveals that many studies deal either with homogeneous systems or treat heterogeneity in an overly corse manner, typically by dividing the populations into just two distinct behavioral groups. 
Our analysis indicates when such simplified modeling of heterogeneity is sufficient and when a more detailed approach is required.

For homogeneous systems, annealed and quenched dynamics are inherently equivalent. 
Furthermore, a heterogeneous system under annealed dynamics can always be mapped to a homogeneous system because its macroscopic behavior depends only on the mean of the preference distribution.
Such systems describe one dimensional flows, so they cannot exhibit oscillations.
In contrast, a heterogeneous system under quenched dynamics may exhibit richer behavior, as it is described by a high-dimensional flow, where oscillations may arise.
This makes it impossible to reduce such a heterogeneous system to a homogeneous one in general.
Only under the balancing condition this reduction is possible. 
Figure~\ref{fig:diagram} summaries these dependencies.
Satisfying the balancing condition has three major consequences \cite{Jed:Men:25}: (1) heterogeneous systems can be mapped into homogeneous ones, (2) annealed and quenched dynamics become equivalent, and (3) any oscillations cannot emerge.

We illustrate these consequences on complex systems across disciplines. In models of social influence, we show that robustness to population heterogeneity is not an intrinsic property of anticonformity itself, but rather a feature of its specific functional representation, and alternative formulations may lead to different conclusions.
Similarly, different social mechanisms can lead to identical update formulas, which underscores the need for caution when using macroscopic patterns to validate microscopic rules \cite{Gal:05,Gal:22}.
This highlights that the implications of different social responses depend crucially on their mathematical formulations.
In models of competing dynamics, we show that the disorder distribution directly impacts the scale of the rounding effect, determining whether a discontinuity is simply weakened or completely eliminated.
Finally, in models of disease spreading, we show that heterogeneity fundamentally alters the dynamics of epidemics. It can accelerate the initial disease spread, but it can also provide a form of collective resistance that reduces the final epidemic size when the pathogen is highly transmissive. 

While our framework provides a tool for evaluating the role of heterogeneity in fully connected graphs, its application to networked populations remains an open challenge.
Additionally, extending this formalism to incorporate multi-flip and multi-state dynamics as well as setups covering multiple competing mechanisms simultaneously, presents a promising avenue to broaden the scope of this unifying theory.


\section*{Acknowledgments}
This work was developed within the scope of the project i3N, UID/50025 and LA/P/0037/2020, financed by national funds through the FCT/MEC. A.J. acknowledges funding from the National Science Centre, Poland under the OPUS call in the Weave programme, project no. 2023/51/I/HS6/02269.

\section*{Data Availability}
The data that support the findings of this article are openly available \cite{Jed:data}.

\appendix
\section{Analytical Calculations for Specific Distributions}
\label{sec:app}

In this section, we present calculations for the populations described by two preference distributions under the quenched dynamics. \ref{sec:one-point} focuses on a degenerate distribution modeling a homogeneous system, whereas \ref{sec:bernoulli} examines a Bernoulli distribution describing a heterogeneous system.

\subsection{Degenerate Distribution}
\label{sec:one-point}
Let $\phi(p)$ be a degenerate distribution (a one-point distribution) with mean $\bar{p}$, i.e.,
\begin{align}
        \phi(p)=&
        \begin{cases}
            1 & \text{if }p=\bar{p},\\
            0 & \text{otherwise}.
             \end{cases}
\end{align}
Such a distribution models a homogeneous population since all the agents have the same preference, $p_i=\bar{p}$.
This modeling approach was used in studies on Ising models \cite{Gar:Lab:Mar:87,Tom:Oli:San:91,Tam:Ale:Gup:94}, or nonlinear voter models \cite{Abr:Paw:Szn:19,Abr:etal:21,Mus:etal:25}.
For homogeneous populations, annealed and quenched dynamics are trivially identical since only one value of preference is possible.
Let us show it explicitly by applying the calculations from Section~\ref{sec:quenched} to the degenerate distribution.
Following the notation presented there, let $a_{\bar{p}}$ denote the fraction of agents in state $A$ with preference $p_i=\bar{p}$.
However, this is also the total fraction of agents in state $A$ in the system, which follows from the discrete verion of Eq.~\eqref{eq:apopulation}:
\begin{equation}
\label{eq:one-point:identity}
    a=a_{\bar{p}}\phi(\bar{p})=a_{\bar{p}}.
\end{equation}
The system evolves according to Eq.~\eqref{eq:rate-que}, which gives
\begin{equation}
    \frac{da_{\bar{p}}}{dt}=P^{\bar{p}}_{BA}(1-a_{\bar{p}})-P^{\bar{p}}_{AB}a_{\bar{p}},
\end{equation}
where the transition probabilities,
\begin{equation}
\begin{split}
    P^{\bar{p}}_{BA}&=\bar{p}X_{BA}(a)+(1-\bar{p})Y_{BA}(a),\\
    P^{\bar{p}}_{AB}&=\bar{p}X_{AB}(a)+(1-\bar{p})Y_{AB}(a),
\end{split}  
\end{equation}
follow from Eq.~\eqref{eq:transition-rates-que}.
Eventually, after combining all the formulas above, we arrive at
\begin{equation}
\begin{split}
    \frac{da}{dt}=&\bar{p}X_{BA}(a)+(1-\bar{p})Y_{BA}(a)-\left[Y_{BA}(a)+Y_{AB}(a)\right]a\\
    &-\left[X_{BA}(a)-Y_{BA}(a)+X_{AB}(a)-Y_{AB}(a)\right]\bar{p}a,
\end{split}
\end{equation}
which is exactly the same equation as in the case of annealed dynamics, Eq.~\eqref{eq:time_evol_ann}, regardless of whether the balancing condition is satisfied or not. 
As the time evolution is the same for both the dynamics, the fixed point must be the same too. 
From Eq.~\eqref{eq:ap_que} and the discrete version of Eq.~\eqref{eq:apopulation-st}, we have
\begin{equation}
    a_{\bar{p}}^*=\frac{Y_{BA}(a^*)-\bar{p}\left[Y_{BA}(a^*)-X_{BA}(a^*)\right]}{Y_{BA}(a^*)+Y_{AB}(a^*)+\bar{p}\left[X_{BA}(a^*)+X_{AB}(a^*)-Y_{BA}(a^*)-Y_{AB}(a^*)\right]},
\end{equation}
where $a^*=a^*_{\bar{p}}$. Finally, after some transformations, we get
\begin{equation}
    \bar{p}=\frac{Y_{BA}(a^*)-a^*\left[Y_{AB}(a^*)+Y_{BA}(a^*)\right]}{Y_{BA}(a^*)-X_{BA}(a^*)+a^*\left[X_{AB}(a^*)+X_{BA}(a^*)-Y_{AB}(a^*)-Y_{BA}(a^*)\right]},
\end{equation}
which is the same as in the annealed case, see Eq.~\eqref{eq:fixed-points-ann}.

\subsection{Bernoulli Distribution}
\label{sec:bernoulli}
Let $\phi(p)$ be a Bernoulli distribution with mean $\bar{p}$, i.e., 
\begin{align}
        \phi(p)=&
        \begin{cases}
            1-\bar{p} & \text{if }p=0,\\
            \bar{p} & \text{if }p=1.
             \end{cases}
\end{align}
In this case, the population consists of two groups of agents as preferences can take only two values, $p_i\in\{0,1\}$. 
Agents with $p_i=0$ always update their states according to mechanism $Y$, while agents with $p_i=1$ always use mechanism $X$.
The fraction of agents in the former group is $1-\bar{p}$ and in the latter $\bar{p}$.
The Bernoulli distribution is commonly employed as a simple distribution that creates a heterogeneous population. It was used in studies on linear \cite{Mas:13} and nonlinear voter models \cite{Jed:Szn:17,Jed:Szn:22, Byr:etal:16, Szn:Szw:Wer:14,Tan:Mas:13}, majority-vote models \cite{Vil:Mor:Sou:12,Vil:Sou:17,Oli:etal:24}, or learning strategies \cite{Yan:etal:21, Jed:Her:24, Wu:etal:25}.

Following the notation introduced in Section~\ref{sec:quenched}, let $a_0$ and $a_1$ be the fractions of agents in state $A$ within the groups characterized by preferences $p_i=0$ and $p_i=1$, respectively. 
Then, the total fraction of agents in state $A$, given by the discrete version of Eq.~\eqref{eq:apopulation}, is
\begin{equation}
    a=a_0\phi(0)+a_1\phi(1)=a_0(1-\bar{p})+a_1\bar{p}.
\end{equation}
The fractions of agents in state $A$ within both groups, $a_0$ and $a_1$, evolve according to Eq.~\eqref{eq:rate-que}, which gives
\begin{align}
    \label{eq:rate-que:bernoulli}
    \frac{da_0}{dt}&=P^0_{BA}(1-a_0)-P^0_{AB}a_0,\\
    \frac{da_1}{dt}&=P^1_{BA}(1-a_1)-P^1_{AB}a_1.
\end{align}
The transition probabilities are obtained by Eq.~\eqref{eq:transition-rates-que}.
For agents with $p_i=0$, we have
\begin{equation}
\begin{split}
\label{eq:transition-rates-que:bernoulli:0}
    P^0_{BA}&=Y_{BA}(a),\\
    P^0_{AB}&=Y_{AB}(a),
\end{split}  
\end{equation}
so we can directly see that this is the group that uses exclusively mechanism $Y$.
For agents with $p_i=1$, we have
\begin{equation}
\begin{split}
\label{eq:transition-rates-que:bernoulli:1}
    P^1_{BA}&=X_{BA}(a),\\
    P^1_{AB}&=X_{AB}(a),
\end{split}  
\end{equation}
so this group uses only mechanism $X$.
Having combined all the above, we arrive at two coupled differential equations,
\begin{align}
    \label{eq:rate-que:bernoulli2}
    \frac{da_0}{dt}&=Y_{BA}(a)(1-a_0)-Y_{AB}(a)a_0,\\
    \frac{da_1}{dt}&=X_{BA}(a)(1-a_1)-X_{AB}(a)a_1,
\end{align}
which govern the time evolution of the system.
Let us show that this quenched dynamics leads to different trajectories that the annealed one.
Using the discrete version of Eq.~\eqref{eq:dadt_que}, we can write the equation for the time evolution of $a$:
\begin{equation}
\begin{split}
    \frac{da}{dt}=&\bar{p}X_{BA}(a)+(1-\bar{p})Y_{BA}(a)-\left[Y_{BA}(a)+Y_{AB}(a)\right]a\\
    &-\left[X_{BA}(a)-Y_{BA}(a)+X_{AB}(a)-Y_{AB}(a)\right]\bar{p}a_1,
\end{split}
\end{equation}
which we can easily compare to Eq.~\eqref{eq:time_evol_ann} for the annealed dynamics. We see that the last term differs as instead of $a$ we have $a_1$ in the quenched case.
However, when the balancing condition holds, the last term disappears, and both dynamics are equivalent.

The fixed points of the dynamics, $\{a_0^*,a_1^*\}$, are given by Eq.~\eqref{eq:ap_que} and the discrete version of Eq.~\eqref{eq:apopulation-st}:
\begin{equation}
    a_0^*=\frac{Y_{BA}(a^*)}{Y_{BA}(a^*)+Y_{AB}(a^*)},
\end{equation}
\begin{equation}
    a_1^*=\frac{X_{BA}(a^*)}{X_{BA}(a^*)+X_{AB}(a^*)},
\end{equation}
where
\begin{equation}
    a^*=a_0^*(1-\bar{p})+a_1^*\bar{p}.
\end{equation}
Combining the above equations, the fixed points satisfy:
\begin{equation}
    \bar{p}=\frac{Y_{BA}(a^*)-a^*\left[Y_{AB}(a^*)+Y_{BA}(a^*)\right]}{Y_{BA}(a^*)\left[X_{AB}(a^*)+X_{BA}(a^*)\right]-X_{BA}(a^*)\left[Y_{AB}(a^*)+Y_{BA}(a^*)\right]}\left[X_{AB}(a^*)+X_{BA}(a^*)\right].
\end{equation}
We can see that this formula is different from Eq.~\eqref{eq:fixed-points-ann}, which describes the fixed points for the annealed dynamics.
Only when the balancing condition holds, the formulas become equivalent.

\bibliographystyle{elsarticle-num} 
\bibliography{main}


\end{document}